# How do the resting EEG preprocessing states affect the outcomes of postprocessing?

Shiang Hu[1], Jie Ruan[1], Juan Hou[2], Pedro Antonio Valdes-Sosa[3,4], Zhao Lv[1*]

[1]Anhui Province Key Laboratory of Multimodal Cognitive Computation, School of Computer Science and Technology, Anhui University, 11 Jiulong Road, Hefei, 230601, China

[2]Department of Psychology, School of Philosophy, Anhui University, Hefei 230601

[3]The Clinical Hospital of Chengdu Brain Science Institute, MOE Key Lab for Neuroinformation, School of Life Science and Technology, University of Electronic Science and Technology of China, Chengdu, China

[4]Cuban Center for Neurocience, La Habana, Cuba

Corresponding author:    shu@ahu.edu.cn (Shiang Hu), kjlz@ahu.edu.cn (Zhao Lv)

# Abstract

Plenty of artifact removal tools and pipelines have been developed to correct the EEG recordings and discover the values below the waveforms. Without visual inspection from the experts, it is susceptible to derive improper preprocessing states, like the insufficient preprocessed EEG (IPE), and the excessive preprocessed EEG (EPE). However, little is known about the impacts of IPE or EPE on the postprocessing in the frequency, spatial and temporal domains, particularly as to the spectra and the functional connectivity (FC) analysis. Here, the clean EEG (CE) was synthesized as the ground truth based on the New-York head model and the multivariate autoregressive model. Later, the IPE and the EPE were simulated by injecting the Gaussian noise and losing the brain activities, respectively. Then, the impacts on postprocessing were quantified by the deviation caused by the IPE or EPE from the CE as to the 4 temporal statistics, the multichannel power, the cross spectra, the dispersion of source imaging, and the properties of scalp EEG network. Lastly, the association analysis was performed between the PaLOSi metric and the varying trends of postprocessing with the evolution of preprocessing states. As results, compared with CE, it was found that: 1) the temporal statistics under IPE and EPE deviated greatly with more noise injected or more brain activities discarded; 2) the IPE power was higher but the EPE power was lower, and the IPE power was almost parallel to that of CE cross all frequencies, while the power difference between the EPE and the CE decreased with higher frequencies; the cross spectra of IPE generally deviated more greatly than the EPE, except for the beta band; 3) derived from the 7 Coupling measures, the IPE network had lower transmission efficiency and worse integration ability, while the EPE network had high transmission efficiency and better integration ability. 4) the source activation under IPE distributed more dispersedly with greater strength while the source under EPE activated more focally with lower amplitude; 5) The PaLOSi was consistently correlated with the varying trends of the investigated postprocessing. This study shed light on how the postprocessing outcomes are affected by the preprocessing states and PaLOSi may be a potential effective quality metric.

# 1. Introduction

The scalp Electroencephalography (EEG) provides a non-invasive recording of the neural activity with millisecond resolution. The spontaneous EEG under the resting state is a powerful tool to explore the brain dynamics, widely applied in the cognitive, neuroscience, psychophysiology, clinical neurophysiology and the brain computer interface (Jas et al. 2017). The weak brain signals detected by electrodes are prone to the contamination of both physiological artifacts such as eye blinks, muscle activations and non-physiological signals such as electrical interference or external environmental factors (Gabard-Durnam et al. 2018). These artifacts may have great impacts on EEG and may eventually lead to spurious conclusions if left without treatment. It is thus indispensable to clean artifacts and correct the recordings during the preprocessing. The EEG/MEG study paradigm is transiting from the single site with small sample to the multiple sites with massive samples, pushing the preprocessing from the manual selection to the automatic correction.

Originating from a single site, the independent component analysis (ICA) is a data-driven approach to replace the human effort and has been the mainstay of artifact removal. It firstly decomposes the recordings into the statistically independent components then performs the artifact subspace reconstruction by discarding the non-brain components. ICA is widely used and proved effective in EEG denoising (Nolan, Whelan, and Reilly 2010) (Jung et al. 2000; Vorobyov and Cichocki 2002). When integrative research evolves the multiple sites and massive sample, the large scale EEG presents the large amount and complex nature (Keil et al. 2014). Various automated preprocessing pipelines have proliferated recently, such as the Harvard EEG Automated Processing Pipeline (HAPPE) (Gabard-Durnam et al. 2018), the Batch EEG Automated Processing Platform (BEAPP) (Levin et al. 2018) and the automatic pre-processing pipeline (APP) (da Cruz et al. 2018), the Computational Test of Automated Preprocessing (CTAP) toolbox (Cowley, Korpela, and Torniainen 2017; Cowley and Korpela 2018), PREP (Bigdely-Shamlo et al. 2015), Automagic (Pedroni, Bahreini, and Langer 2018), EEG Integrated Platform Lossless pre-processing pipeline (EEG-IP-L) (Desjardins et al. 2021). These pipelines support the DC removal, filtering, interpolation, the referencing, the ICA, and other procedures and lastly provide quality metrics for quantitative evaluation.

All the artifact correction and the preprocessing pipelines boil down to the point of quality control. The essence of quality control is not to pursue the normal EEG waveforms but to have objective and lossless preprocessing with the goal that the preprocessing has no bias on the postprocessing outcomes. In practice, it is hard to achieve complete removal of artifacts and keep brain signals intact since the clean EEG is never known. Alternatively, the preprocessing can be ascribed into proper preprocessing and improper preprocessing. However, currently little is known about the impacts of the resting EEG preprocessing states on the postprocessing outcome (Kaiser et al. 2021; Pellegrino et al. 2022), hindering the EEG quality evaluation and assurance.

How to ensure maximum noise removal and retention of EEG signals is a problem to be addressed by quality control (Desjardins et al. 2021). Since the discovery of EEG, a rich family of postprocessing methods with neurophysiological interpretations has been formed, which can be categorized into temporal, frequency, and spatial domain analysis. As to the resting EEG, the temporal domain analysis such as the amplitude, variance, the field power, can only capture the shallow statistics of the brain waves, whereas the frequency or spatial domain analysis, such as power spectra, cross-spectra, source reconstruction, functional connectivity, EEG brain network provide the depth understanding of brain dynamics.

This study aims to investigate how the EEG preprocessing affects the postprocessing outcomes in the temporal, the frequency, and the spatial domains. The EEG postprocessing methods may vary from lab to lab, depending on the research purpose. Besides the group statistics and the pattern learning such as the classification, the regression, and the casual inference, the core of postprocessing after the artifacts cleaning and correction are the feature engineering, e.g., the EEG biomarker identification for psychophysiological studies. Constraint to the quantitative EEG field, the postprocessing are mainly the power analysis over the broadbands, the frequency functional connectivity analysis, the source analysis, and the EEG network analysis. As demonstrated in (Nolte et al. 2020), the cross-spectrum determines the statistical properties of the asymptotic Gaussian distributed data, and all normalized coupling measures in the frequency domain are the simple close functions of complex coherency. Thus, this study focuses on the investigation that how the preprocessing states affect the profiles of cross spectra tensor and the fidelity of functional connectivity. Meanwhile, the source imaging and the temporal statistics are also analyzed to cover as more outcomes in the EEG postprocessing.

EEG potentials originate from the macroscale neural oscillations and consists of a series of rhythmic activities. Spectral analysis converts the EEG signal from the time domain to the frequency domain by means of the Fourier transform and describes the distribution of Spectral analysis, the energy or power of the EEG signal in frequency (Zhang 2019). The EEG functional connectivity (FC) matrices can be drawn from the coupling measures in the temporal or frequency domains. EEG FC describes the pairwise association between the activities of any two electrodes. The common coupling measures to build FC are the coherence (Coh), the imagery part of the coherence (iCoh), the phase lag index (PLI), the phase locking value (PLV), the Ganger causality (GC), the partial directed coherence (PDC), the directed transfer function (DTF). Although the sensor FC does not represent the true interactions of brain regions underneath the electrodes due to the volume conduction, the scalp EEG FC is still important considering that it is the intermediary step to estimate the true source FC and directly used for the application such as brain age prediction, emotion recognition and brain states classification. The impact of noise on spectra and brain networks has been reported. Specifically, Bastos and Schoffelen 2016 demonstrated that adding white noise to a channel can cause an upward shift in the power spectrum. Both Bastos and Schoffelen 2016 and Pellegrino et al. 2022 indicated that noise can weaken network connections, leading to distortion in the network. Additionally, Yu 2020 conducted experiments to assess the influence of noise on networks constructed using different methods.

Besides, a typical spatial analysis is source imaging. The ill-posed nature of source imaging leads to the non-uniqueness of inverse solution. The classical inverse solutions are the nonparametric optimization methods such as the minimum norm estimation (MNE), the LORETA modified variants, and VARETA and the parametric methods such as the beamforming and the BESA and the MUSIC method (Grech et al. 2008). The measures to evaluate the distributed source imaging are the dipole localization error, the source dispersion, and the resolution index (Molins et al. 2008) which are all resolution matrix metrics.

This paper is organized as follows: In the "**Materials and Methods**" section, we firstly show how the different preprocessing states were synthesized, then presented how the impacts of preprocessing EEG on the spatial domain, the frequency domain, and the time domain were sequentially explored, and thirdly proposed the PaLOSi as a potential quality control metric and performed the association analysis between PaLOSi with varying trends of the preprocessing outcomes. The "**Results**" section described how the postprocessing outcomes were affected by the preprocessed EEG quality. The rest of this paper is the detailed "**Discussion**" and a brief conclusion.

## 2. Materials and methods

### 2.1 Simulation

During the online recording, the purely clean EEG (CE) is hardly known in this raw acquisition step but is still an ideal state pursed by the data analysts. The offline preprocessing operations to obtain CE is recognizing the noise and then rejecting the bad epochs/components from the raw recording by advanced tools or the experts' selection. However, the extent of preprocessing lacks objective standard. In contrast, a common state is the insufficiently preprocessed EEG (IPE), that is, the preprocessing did not completely clean the noise. Another easily neglected state is the excessively preprocessed EEG (EPE), that is, the brain activities were unintentionally discarded in denoising, especially when powerful statistical methods were applied, such as PCA, ICA, bad channel removal, etc. Thus, CE, IPE and EPE are the three states to be investigated in this study.

As a benchmark for quality assessment and validation of connectivity estimation, the SEED-G allows for mimicking the real EEG properties, such as the non-stationary, the normative EEG spectral properties, the intertrial variability, and the time varying networks. The SEED-G offers flexibility in adjusting the time series properties and forming the source connectivity patterns (Anzolin et al. 2021). The noise-free EEG was synthesized by the volume conduction model only without considering sensor noise as the CE. Noises with varying intensities were combined with CE to mimic the IPE, and brain components with varying explained variances were discarded to mimic the EPE. **Fig. 1** provides an overview of the main workflows and the simulated EEG data generation process, including the definition of the ground truth, IPE, and EPE.

### 2.1.1 Generation of clean EEG as benchmark

With SEED-G, we defined the brain sources by picking the cortical regions of interest (ROIs) from the Broadman area. The cortical parcellation of ICBM152 New York anatomy template comprised of 15002 vertices and used the Broadman areas that consists of 79 brain regions. The sensor space was configured with 15 channels covering the whole brain. The lead field matrix with the size of 15 by 15002 was computed by the OpenMEEG in Brainstorm toolbox given the parcellations by Freesurfer (Haufe and Ewald 2019). Only 15 cortical regions of interest (ROIs) were picked from the Broadman areas as the active brain regions. The 10-order multivariate autoregressive (MVAR) model with the size 15x15x10 were taken as the generating filter of source activities. The MVAR model in each lag is a weighted and directed matrix. 5% of the MVAR coefficients except for the diagonals were randomly imposed within the range [-0.5, 0.5] as the non-null connections. To ensure that the simulated EEG can maintain the real EEG spectral properties, the 50% ROIs plus the isolated ROIs were assigned the AR coefficients estimated from the source activities using the sLORETA on a resting EEG from a healthy subject.

Then, the ROI times series were generated by the MVAR model and the innovation process with identity covariance. The ROI activities were mixed with Gaussian noise with the predefined SNR 10dB. Thus, the source signals consisting of 15 ROIs were generated and considered as the ground truth (GT) source activity. The dipolar sources in each ROI are set to have identical activities. The noise-free EEG was synthesized by the volume conduction model without considering sensor noise as the clean EEG (CE). Finally, the CE was synthesized with the size 15 channels and 2500-time samples. For the statistical analysis, the same simulation procedure was repeated 200 times, in each time of which the locations and the values of the non-null connections in the MVAR model, the brain sources that were assigned the source activity estimated from the real EEG, and the Gaussian noise in the source space were varied.

The general form of MVAR model can be defined as:

$$\mathbf{j}(t) = \sum_{i=1}^{p} \mathbf{A}(i)\mathbf{j}(t-i) + \xi(t) \tag{1}$$

$$\mathbf{v}_c = \mathbf{K}\mathbf{j} \tag{2}$$

where $\mathbf{A} \in \mathbb{R}^{N_k \times N_k \times p}$ stores the MVAR model coefficients, $\mathbf{j}$, $\xi \in \mathbb{R}^{N_k \times 1}$ are the source activity and the innovation noise, $\mathbf{K} \in \mathbb{R}^{N_c \times N_k}$ is the lead field matrix, $\mathbf{v}_c \in \mathbb{R}^{N_c \times 1}$ is the instantaneous CE potentials, $N_c$, $N_k$ and $p$ denote the number of the channels, the number of selected ROIs, and the MVAR model order, respectively. The innovation noise was generated from a multivariate normal distribution with zero mean and identity covariance.

### 2.1.2 Practical preprocessing EEG states

**(1) Insufficiently preprocessed EEG (IPE)**

The preprocessing has no objective standard and consistent form, which ranges from the manual rejection of bad epochs, the semi-automatic or the semi-supervised denoising by the hybrid use of mini algorithms and toolkits, to the fully automatic preprocessing pipelines with user friendly toolbox or adapted for the terminal parallel processing. Hence, the extent of EEG preprocessing is hard to quantify and compare. The IPE is generated by introducing sensor noise to the CE as follows:

$$\mathbf{v}_i = \mathbf{v}_c + \boldsymbol{\varepsilon} \tag{3}$$

$$r_n = 10\log_{10}(\|\mathbf{v}_i\|^2 / \|\boldsymbol{\varepsilon}\|^2) \tag{4}$$

where $\mathbf{v}_i$ denotes the IPE. $\boldsymbol{\varepsilon}$ is the spatially and temporally uncorrelated Gaussian noise. $r_n$ is the signal to noise ratio (SNR). The smaller $r_n$ implies the more insufficient preprocessing (MIP).

**(2) Excessively preprocessed EEG (EPE)**

It is intuitively supposed that all the ICs separated from the CE are "brain ICs". However, after performing ICA on the CE, all the separated ICs may be still labeled as the brain ICs and the non-brain ICs such as the channel noise, the muscle activity, etc. The CE is expressed as

$$\mathbf{v}_c = \mathbf{As} = [\mathbf{A}_{rt}, \mathbf{A}_{rj}]\left[\mathbf{s}_{rt}^T, \mathbf{s}_{rj}^T\right]^T \tag{5}$$

where $\mathbf{A}$ is the IC mixing matrix consisting of the weights of component maps in the columns, which is partitioned into the $\mathbf{A}_{rj}$ to be rejected and the $\mathbf{A}_{rt}$ to be retained. $\mathbf{s}$ are the instant IC activations, which is partitioned as the $\mathbf{s}_{rj}$ to be rejected and the $\mathbf{s}_{rt}$ to be retained. The back projection from the rejected IC activations to the EEG time series is as

$$\mathbf{v}_{rj} = \mathbf{A}_{rj}\mathbf{s}_{rj} \tag{6}$$

The ICs were sorted from smallest to largest according to the percentage of variance accounted for (PVAF) the recording, and then the selected components were removed in combination. The signal distortion ratio (SDR) $r_l$ was defined based on the PVAF of the removed components, with $Var$ denotes the variance,

$$r_l = 1 - \mathbf{1}^T Var(\mathbf{v}_{rt}) \big/ \mathbf{1}^T Var(\mathbf{v}_c) \tag{7}$$

The EPE with the SDR $r_l$ is synthesized as:

$$\mathbf{v}_e = \mathbf{A}_{rt}\mathbf{s}_{rt} \tag{8}$$

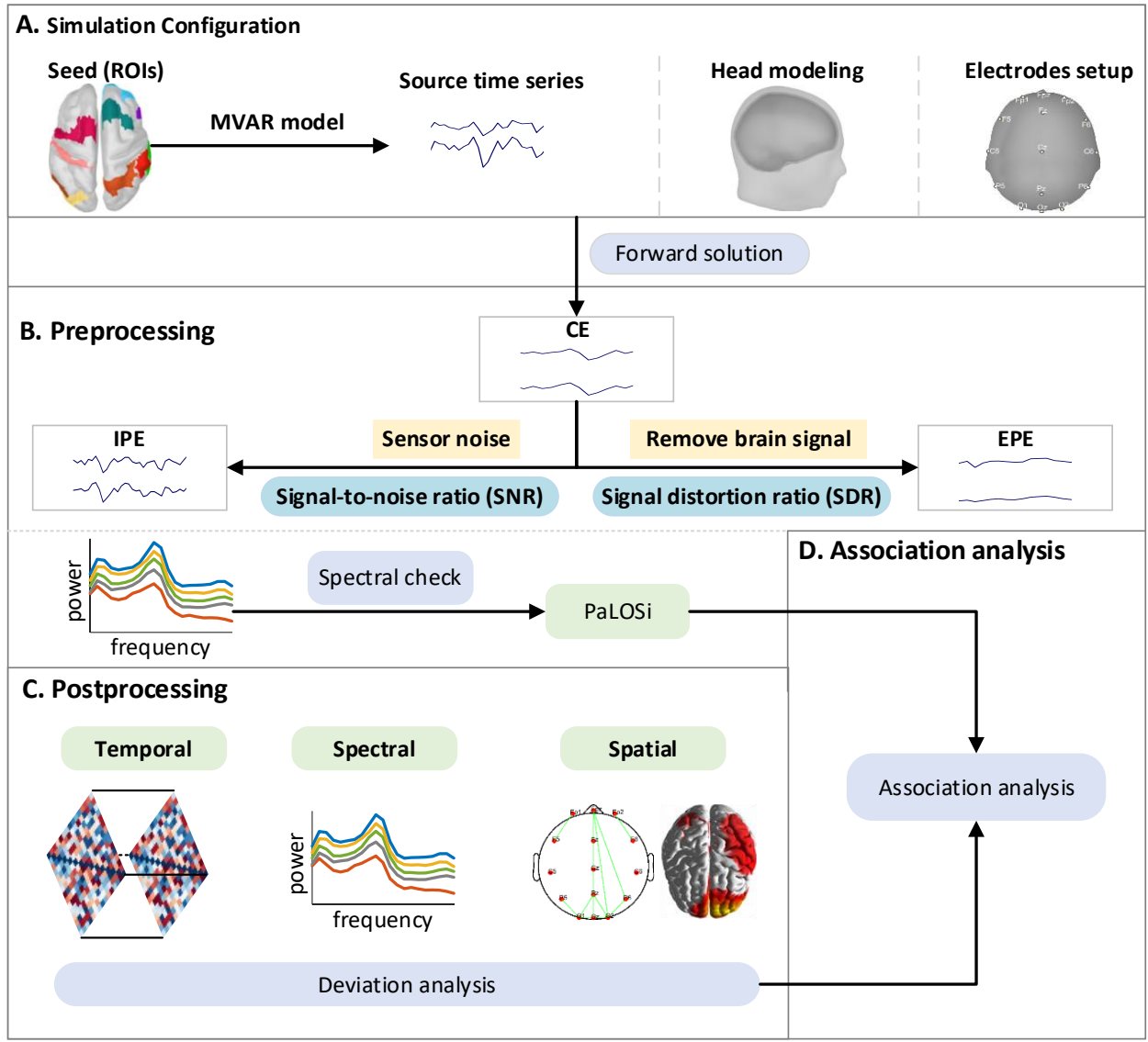

Fig. 1. Schematic flowchart. A. The source, electrodes, and head model configuration for forward solver; B. The synthesis of CE, IPE, and EPE; C. Postprocessing as to the source imaging, the network, the spectrum, and the temporal aspects.

The simulation scheme was designed to control the extent of preprocessing, which results in multiple IPEs, CE, and multiple EPEs. The extent of IPEs and the EPEs was tuned by different SNRs and SDRs, correspondingly. The three classes of simulated EEGs under different preprocessing extents parallelly undergone the downstream postprocessing analysis from the temporal, frequency, and the spatial domains.

## 2.2 Temporal analysis

### 2.2.1 Temporal measures

Although the temporal analysis (TA) is not much involved in the quantitative EEG field, it remains the nonnegligible necessities for the time series modeling, the waveform statistics, and the temporal information decoding in various fields such as the brain computer interface, the EEG emotion recognition, the nonlinear neural dynamics, and the artifact removal. The TA typically are the mean, the standard deviation, the zero crossings, the cross correlations, the covariance, the complexity, and the entropy. However, compared to the analysis that is strictly ascribed to the temporal domain, the TA are likely analyzed with other domains, in the way of the hybrid analysis across domains, such as the time-frequency and the spatiotemporal analysis.

We here therefore focused on the multivariate statistics derived from the time domain to investigate the impacts of preprocessing on spatiotemporal analysis. The investigated measures were the cross-channel Pearson correlation, the cross-channel covariance, and the amplitude-amplitude coupling, that is, the amplitude envelope correlation (AEC) (Bruns et al. 2000), and the power envelope correlation (PEC) with correction for artifacts of volume conduction (Hipp et al. 2012). The AEC estimates the coupling by the correlation between the low frequency amplitude and the high frequency amplitude. The PEC is the Pearson correlation between the power envelopes after taking the absolute values of the log transformed complex spectra. The multivariate analysis forms the graphs under different preprocessing states such as CE, IPE, and EPE.

### 2.2.2 Deviation analysis

In the deviation analysis, the Riemannian distance-based graph comparison algorithms (Cordella et al. 2004) was applied to characterize the similarity between graphs where regularization may be used as the equation (9). The significance lies in analyzing how the preprocessed EEG quality affects the statistical similarity from the temporal outcomes from the CE.

## 2.3 Spectral analysis

### 2.3.1 Spectral estimation

The typical EEG spectral analysis includes the multichannel power spectra estimation, the cross-spectra estimation, the time-frequency analysis, and the high-order spectral analysis. Spectral estimation captures time-invariant profiles of the quasi-stationary epochs. The multichannel power spectra and the cross spectra are the two crucial estimators for the quantitative EEG analysis. Especially, the cross spectra can derive the coherence by taking normalization. The power spectra of IPE, CE, and EPE in the range of 1-40Hz, and the cross-spectra in the delta (1-4 Hz), theta (5-8 Hz), alpha (9-13 Hz), and beta (14-30 Hz) bands were estimated.

### 2.3.2 Deviation analysis

In the deviation analysis, for the power spectrum, the Euclidean distance of power spectra under the improper preprocessing (IPE, EPE) from that under the CE were calculated for statistical differences. Since the cross-spectra are embedded in the Riemannian space, the deviation of cross-spectra under IPE and EPE from that of the CE is measured using the Riemannian distance using the Riemannian metric. The cross-spectral deviation at each frequency band is computed by taking the mean of Riemannian distances over the multiple frequencies distributed in the investigated frequency bands.

To ensure that the cross-spectra is positive definite, the regularization was implemented as (Li et al. 2022),

$$\begin{cases} \tau = |\min\{\boldsymbol{\lambda}\}| + 1 \\ \hat{\mathbf{S}} = \dfrac{1}{1+\tau}\mathbf{S} + \dfrac{\tau}{1+\tau}\mathbf{I}_{N_e} \end{cases} \quad (9)$$

where $\mathbf{S}$ and $\hat{\mathbf{S}}$ denote the original cross-spectral matrices and the regularized cross-spectral matrices, $\boldsymbol{\lambda}$ is the vector stacking the eigenvalues of $\mathbf{S}$.

## 2.4 Scalp EEG network

### 2.4.1 Coupling measures

The functional connectivity (FC) estimation is a prerequisite step to build the EEG network. The FC value refers to the pairwise statistical correlation or dependency derived from the information theory. Here, seven Coupling measures were adopted, such as the coherence (Coh), the imaginary part of coherency (iCoh) (Nolte et al. 2004), the partial directed coherence (PDC) (Baccalá and Sameshima 2001), the directional transfer function (DTF) (Baccalá and Sameshima 2001), the Granger causality (GC) (Brovelli et al. 2004), the phase-locked value (PLV) (Lachaux et al. 1999) and the phase slope index (PSI) (Nolte et al. 2008). Note that although the amplitude envelope correlation (AEC) and the power envelope correlation (PEC) can be used as Coupling measures, both were studied in the temporal analysis.

Coherency is the normalized cross spectrum and coherence is the absolute value of coherency. The iCoh is helpful to estimate the true interactions between the brain sources since it is insensitive to the instantaneous volume conduction effect. As a classical measure of phase-phase coupling, the PLV gives the modulus of the averaged instantaneous phase differences between two time series. The PSI estimates the direction of information flow, based on the slope of the phase difference between two signals. The GC measures how the future of X can be predicted when the past of Y is included. Both PDC and DTF give the casual relation between the outflow of node X towards Y in the frequency domain based on the Granger causality principle but differs by normalizing by all the outflows from the node X and by all the inflows towards node Y.

### 2.4.2 Network metrics

With the electrode sites as the nodes and the FC values as the edge weights, the EEG networks under the CE, IPE, and EPE states were constructed from the FC matrices. In the network neuroscience field, rich metrics have been proposed to characterize the properties of brain network, such as degree, degree distribution (Barabási and Albert 1999), shortest path length, clustering coefficient (Watts and Strogatz 1998), global efficiency, local efficiency (Latora and Marchiori 2001), participation coefficient (Guimerà and Amaral 2005), assortative coefficient (Newman 2002), number of triangles, transitivity (Newman 2003), modularity (Newman 2004), closeness centrality, betweenness centrality (Freeman 1978), small-worldness (Humphries and Gurney 2008).

Here, four widely used metrics were selected as the representative ones for the deviation analysis, which are the characteristic path length (CPL), the clustering coefficient (CC), the global efficiency (GE), and the local efficiency (CE). Its interpretations are as follows:

1) CPL: the average of shortest path length between all node pairs. The smaller CPL indicates that the information flow between nodes costs less in shorter time.
2) CC: the ratio of the weighted geometric mean of triangles in the current network to the maximum number of possible edges. A higher CC indicates a more efficient and robust interaction in the network, a higher level of local clustering or community structure, while a lower CC suggests a more dispersed or random network configuration.
3) GE: the inverse of average shortest path length. The GE quantifies the transmission efficiency of the whole network. The higher GE means the faster information flow and the greater network integration ability (Ismail and Karwowski 2020).
4) LE: the inverse of the shortest path length that only passes the current node. It quantifies the ability to integrate information in the local state of the network.

### 2.4.3 Deviation analysis

Graph theory is a powerful tool for understanding, characterizing, and quantifying complex brain networks (Huang et al. 2018). Several methods for quantifying the graph similarity have emerged, such as the graph edit distance, the DelataCon method, and the graph kernel approach. SimiNet is a novel method designed to quantify the brain network similarity that accounts for node, edge and spatiality features (Mheich, Wendling, and Hassan 2020; Mheich et al. 2018). Based on the SimiNet, we proposed the distance estimator to measure the similarity between the two adjacency matrices, that is, the degree of deviation between two EEG networks. Given two weighted adjacency matrices $\mathbf{C}^1$ and $\mathbf{C}^2$ with the same nodes, the SimiNet distance between the two matrices is defined as follows,

$$d = 1 - 1 \Big/ \Big(1 + \sum_{k=1}^{N_c-1} \sum_{p=k+1}^{N_c} \left| c_{p,k}^1 - c_{p,k}^2 \right| \Big) \tag{10}$$

where $c_{p,k}$ denotes the weight of edge connecting the nodes $p$ and $k$ in the functional connectivity matrix. Here, the SimiNet distance $d \in [0,1]$, 0 and 1 indicate the two connectivity matrices are identical and completely dissimilar, respectively. After the EEG networks under CE, IPE and EPE were constructed, the distance ($d_{(i,c)}$, $d_{(e,c)}$) of improper preprocessing (IPE, EPE) was calculated against the benchmark CE state.

Besides, the deviation of network properties was analyzed by measuring the distance of network metrics between the IPE, EPE states and the CE state.

## 2.5 Source analysis

### 2.5.1 Source reconstruction

The distributed source imaging in the frequency domain was solved by using eLoreta (Pascual-Marqui et al. 2011). Using the ICBM152 anatomical template, the head model was computed with the OpenEEG in Brainstorm and exported in the corresponding FieldTrip format. By means of the Fieldtrip toolbox, the lead field was estimated based on the electrode registration, the head model and the source model (Oostenveld et al. 2011). The bandlimited source activation in each frequency band was the mean of the activated strength over all the frequencies within the $\sigma$ band(0.5-4 Hz), the $\theta$ band (4-8 Hz), the $\alpha$ band (8-13 Hz), and the $\beta$ band (20-25 Hz).

### 2.5.2 Deviation analysis

The impacts of preprocessing on the source localizations were evaluated by the dispersion extent of activated source as well as the activated strength. Note that the dipole localization error (DLE) and the resolution index (RI) are the two popular metrics to evaluate the source imaging methods. Here, the attention to source imaging is how dispersedly the sources are distributed due to the IPE and especially the EPE. Based on (Molins et al. 2008), the spatial dispersion (SD) was defined as:

$$\text{SD}_i = \sqrt{\sum_j d_{ij} \mathbf{R}_{ij}^2 \Big/ \sum_j \mathbf{R}_{ij}^2} \tag{11}$$

where $d_{ij}$ is the distance between source j and i, and $\mathbf{R}$ is the resolution matrix of the linear inverse estimator, the columns and the rows of $\mathbf{R}$ are the point spreading functions (PSF) and the cross-talk functions (CTF), respectively. The PSF represents the estimated source distribution for a single point source and the CTF describes the sensitivity of an amplitude estimator for a single point source to all other sources (Hauk, Wakeman, and Henson 2011). The SD value can reflect both displacement in PSF from the true source distribution and a spatial broadening of the PSF. The larger SD value indicates the more widely distributed PSF. Physically, the SD may in a large extent reflect how the distributed source imaging can be ascribed to the dipole source imaging and, or say, the main components and the homogeneity in the source space.

## 2.6 Association analysis

### 2.6.1 PaLOS index (PaLOSi)

The deviation of postprocessing outcomes from that of CE are intuitively caused by the improper preprocessing, that is, the preprocessed EEG quality. Thus, alleviating the distortion of postprocessing boils down to the EEG quality control which is the core concern of this exploratory study. The key to this issue is to develop a quality metric that is sensitive to the varying trends of postprocessing deviation. However, most of the current existing quality metrics are the EEG waveform indices in the temporal aspects, such as the ratio of bad channels, the ratio of data with high amplitude, the ratio of timepoints of high variance, ratio of the channels with high variance. Here, we proposed the Parallel Log Spectra index (PaLOSi) as a novel EEG quality indicator by checking the structural homogeneity of cross spectral matrices across frequencies.

In the frequency-wise view, the cross-spectral matrix is the sample covariance of multichannel Fourier series across segments (Brillinger 2001). The multichannel Fourier series subscripting by the frequency $\omega$ is expressed as $\mathbf{\Phi}_\omega \in \mathbb{C}^{N_c \times N_s}$, with $N_s$ as the number of segments. Assuming the spatial whitening matrix is $\mathbf{\Gamma}^\dagger$ and the diagonal matrix is $\mathbf{\Lambda}_\omega$, then the principal component decomposition on the multichannel Fourier series is

$$\mathbf{\Psi}_\omega = \mathbf{\Gamma}_\omega^\dagger \mathbf{\Phi}_\omega \tag{12}$$

$$\mathbf{\Gamma}_\omega^\dagger \mathbf{\Phi}_\omega \mathbf{\Phi}_\omega^\dagger \mathbf{\Gamma}_\omega = \mathbf{\Lambda}_\omega \tag{13}$$

$$\mathbf{S}_\omega \propto \mathbf{\Phi}_\omega \mathbf{\Phi}_\omega^\dagger = \mathbf{\Gamma} \mathbf{\Lambda}_\omega \mathbf{\Gamma}^\dagger \tag{14}$$

The presence of the cross-spectral homogeneity across frequencies can be attributed by both the prominent dominance of the largest eigenvalue and the frequency-invariant property of the eigenvectors. The largest eigenvalue corresponds to the largest explained variance in the principal component space. The whitening matrix that comprised of eigenvectors is identical to the principal decomposition at all the frequencies. This ensures that the cross spectral matrices at each frequency can be mapped into the common orthogonal coordinate space to disclose the cross spectral homogeneity. The spatial transformation applied to the Fourier series is identical across frequencies. The stepwise common principal component (CPC) decomposition method (Trendafilov 2010) can reduce the dimensions and sort the eigenvectors according to the explained variance simultaneously. The stepwise CPC analysis is applied on the cross spectra as

$$\mathbf{S}_\omega = \mathbf{\Gamma} \mathbf{D}_\omega \mathbf{\Gamma}^\dagger \tag{15}$$

and output the matrix $\mathbf{\Gamma}$ and the diagonal matrix $\mathbf{D}_\omega$. The PaLOS index (PaLOSi) is defined as

$$PaLOSi = \sum_\omega \max\{diag(\mathbf{D}_\omega)\} \Big/ \sum_\omega tr(\mathbf{S}_\omega) \tag{16}$$

Here, $\max\{diag(\cdot)\}$ is to extract the largest entry in the main dignoals, and $tr(\cdot)$ is the matrix trace operator. Thus, the PaLOSi lays in the range of [0, 1]. The PaLOSi is the proportion of the variance explained by the first common principal component of the cross-spectral tensor. The larger PaLOSi means that the frequency coupling profiles stored in the cross-spectral tensor are more likely to be homogenous across frequencies.

### 2.6.2 Correlation between the deviations and the PaLOSi

After the repetitive simulation of 200 times, the EEG dataset with the dimension of 64 channels by 2500-time samples by 9 states by 200 repetitions were synthesized. The 9 states include the IPE with 4 different SNRs, the CE, and the EPE with 4 different SDRs. The large datasets allow for the statistical mapping of how the postprocessing analysis varies with the preprocessing states. Here, the investigated postprocessing measures are summarized in the Table 1.

*Table 1. The summary of postprocessing measures in the temporal, the frequency, the spatial aspects.*

| Domains | Postprocessing measures | Deviation analysis |
|---|---|---|
| Temporal | Covariance<br>Correlation<br>Amplitude envelope correlation<br>Power envelope correlation | Riemannian distance |
| Frequency | Power spectra | Euclidean distance |
| | Cross spectra | Riemannian distance |
| Spatial | Scalp EEG functional connectivity | SimiNet based distance |
| | Scalp EEG network metrics: Characteristic path length, Clustering coefficient, Global efficiency, Local efficiency | Euclidean distance |
| | Source imaging | Spatial dispersion |

The downstream analysis was systemically quantified by a couple of measures from the temporal, the frequency, and the spatial aspects. The deviation of these measures reflected how far the postprocessing outcomes were from the ground truth CE by tuning the preprocessing states from IPE to EPE. Since the magnitudes of all the postprocessing measures were in different scales, their deviations were relatively normalized prior to performing the association analysis. The deviations of a postprocessing measure under all the preprocessing states were normalized as

$$\hat{\boldsymbol{\mu}} = \frac{|\boldsymbol{\mu}| - |\mu|_{\min}}{|\mu|_{\max} - |\mu|_{\min}} \qquad (17)$$

where $\boldsymbol{\mu}$ stands for a vector consisting of the measured deviations under the preprocessing states from that under the CE, $|\cdot|$ denotes the absolute value operator, $|\mu|_{\max}$ and $|\mu|_{\min}$ are the maximum and the minimum absolute deviation among all the preprocessing states.

The association analysis was performed between the normalized quantification measures and the PaLOSi. We calculated the PaLOSi of three types of EEG from 200 simulation data, with the aim to learn the trends of PaLOSi varying with the EEG quality and further verify the reliability of PaLOSi as an EEG quality indicator. Note that the deviation of PaLOSi was normalized in the same way. The correlation between the PaLOSi and the varying trends of postprocessing measures aimed to discover the ability of PaLOSi to represent EEG quality.

## 3. Results

### 3.1 Impacts on the temporal measures

The **Fig. 2** shows how the temporal analysis of different preprocessing states deviated from the CE in a 3D plot. The x, y, z axes of the 3D plot represent the temporal measures, the preprocessing states, and the temporal deviation (TD) quantified by the Riemannian distance, respectively. In the 3D graph, warm colors represent IPE, while cool colors represent EPE. *dB represents the magnitude of the signal-to-noise ratio, and *% indicates the degree of loss in brain signals.

It is apparently shown that the COV derived the smallest TDs among the 4 investigated temporal measures and the TDs of COV are consistently smaller than 0.2 as to the preprocessing states. The TDs of the temporal measures except for the COV were considerably larger and changed more dramatically than that of COV.

In addition, the TDs of each temporal measure consistently followed the increasing trend with either the SNR decreasing as to IPE or the SDR increasing as to EPE. This implies that the temporal analysis deviated more from the CE, if the case either the more insufficient denoising or more brain activities being lost happened.

Besides, for the temporal measures except for the COV, their TDs under EPE with a fixed SDR follow the order COR<AEC<PEC, while the TDs under IPE have no clear relation. It can be inferred from all the above that preprocessing states have impacts on the temporal analysis. Especially, the impacts on the functional relations based on the temporal domain information cannot be negligible, whereas the covariance analysis may not obviously suffer from the improper preprocessing.

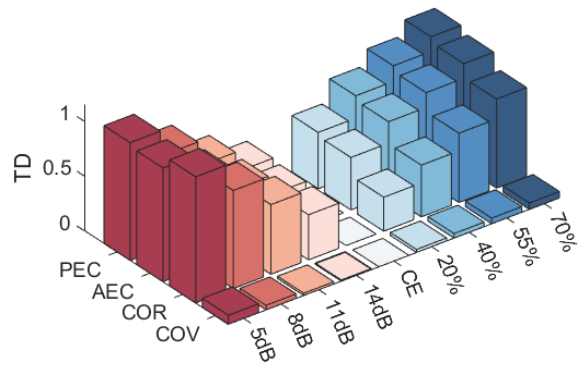

*Fig. 2. The temporal deviations (TDs) under the improper preprocessing states. The TDs of COV, COR, AEC, and PEC under the 9 preprocessing states from the CE. COV: covariance, COR: Correlation, AEC: amplitude envelope correlation, PEC: power envelope correlation.*

### 3.2 Impacts on the frequency spectra

To investigate the effect of the preprocessed EEG quality on the frequency spectrum, we calculated the power deviation (PD) of IPE and EPE from CE in the range of 1-40 Hz and the cross spectral deviation (CD) by the Riemannian distance under the IPE and EPE states from that under the CE state in the conventional four broad bands, that is, the delta (1-3 Hz) band, the theta (4-7 Hz) band, the alpha (8-13 Hz) band and the beta (14-30 Hz) band.

    The **Fig. 3**A shows the PD in the line chart with the x axis in the range of 1-40Hz and the y axis in the dB scale. The multiple curves represented how the power under a specific preprocessing state deviated from the CE state across the frequencies. The different colors that differentiate the curves correspond to the 9 preprocessing states shown in the legend at the bottom. The very light gray color as a flat line at zero intuitively means that the power of CE had no deviation from itself. The positive and the negative PD in dB scale represents greater and lower power than the power of CE. The IPE in warm color had higher power with the lower SNR, while EPE in cold color had lower power with the higher SDR. Note that the curves in warm color are nearly parallel and flat lines, meaning that the PD caused by IPE was almost the same across all frequencies. By contrast, the curves in cold color from the low to the high frequencies are increasing toward the zero flat line, meaning that the PD caused by EPE decreased with the frequency increasing.

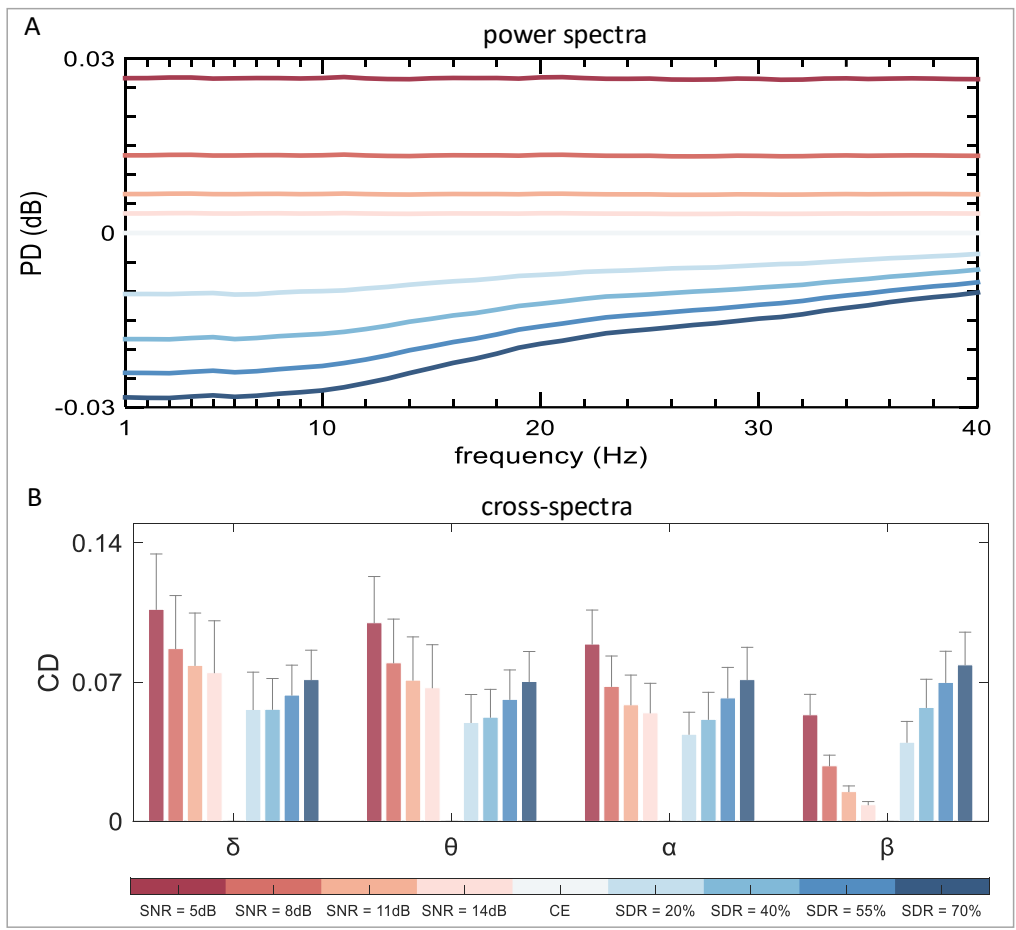

*Fig. 3. **Impacts of EEG preprocessing on the spectral analysis.** **A**. the power deviation (PD) of preprocessing states compared to CE, **B**. the cross spectral deviation (CD) of preprocessing states compared to CE at 4 frequency bands. The legend: The IPE and the EPE are separated in warm color and the cold color, respectively.*

The **Fig. 3**B depicted the cross spectral deviation (CD) of the preprocessing states in the bar plots with the frequency bands in the x axis and the CD in the y axis. The different colors to separate the bars correspond to the 9 preprocessing states shown in legend at the bottom. The CD in each bar was the averaged Riemannian distance over the frequencies within that broad band between the cross spectra under a preprocessing state and that under CE.

    It was consistent for all the frequency bands that the CD increased with more insufficient preprocessing with the SNR decreasing and more excessive preprocessing with the SDR increasing. This reveals that the cross spectra deviated more from the ground truth CE with either more sensor noise introduced or with more brain signals lost.

    From the delta to the alpha band, the CDs of IPE in warm color were generally higher than that of EPE in cold color, while in the beta band the CD of IPE in warm color were generally lower than that of EPE in cold color. This may suggest that IPE caused greater CD than EPE in the low frequency range covering the delta, theta, and alpha bands, while EPE caused

more CD in the high frequency range that is the beta band. In addition, the CD caused by IPE generally reduced with frequency bands shifting from the low to the high frequencies, while the CD caused by EPE showed no prominent difference among frequency bands.

## 3.3 Impacts on the scalp EEG network

Here we investigated both the structural pattern and the graph theoretic metrics of scalp EEG networks constructed by the 7 coupling measures under the 9 preprocessing states. The 7 coupling measures were COH, iCOH, DTF, GC, PDC, PLV, and PSI. The characteristic path length (CPL), the clustering coefficient (CC), the global efficiency (GE), and the local efficiency (LE) were calculated with the IPEs, CEs, and EPEs.

### 3.3.1 Fidelity of the connectivity matrices

**Fig. 4** showed the deviation of the FC matrices constructed by the PLV under improper preprocessing (IPE, EPE) from that under the CE state. The deviation was quantified by taking the mean SimiNet-based distances across all the frequencies between the FC matrices of the EEG under a particular preprocessing state and that under the CE state.

It is easily found that with more insufficient (MIP) or excessive (MEP) preprocessing, the DEVI of FC under either IPE or EPE became greater. However, by contrast, the boxes in cold color were more clustered, while the boxes in warm color were more dispersed. This suggested that the redundant sensor noise will easily distort the FC patterns; minor sensor noise may not contaminate the FC pattern, while the loss of brain signals will easily distort the FC patterns.

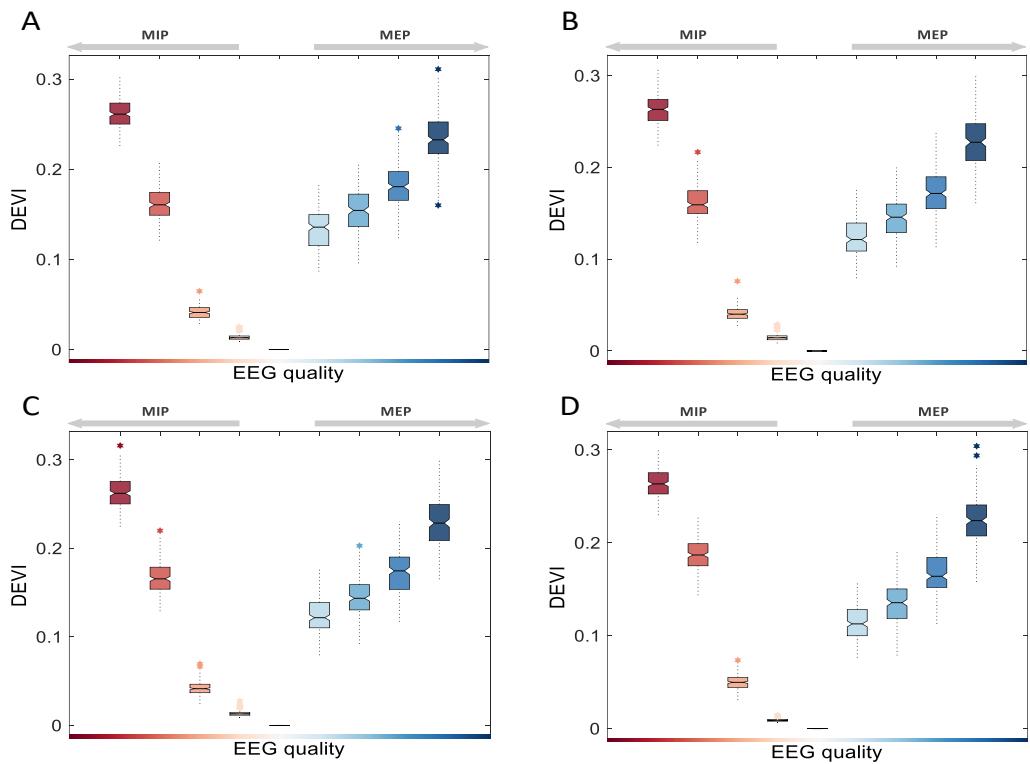

*Fig. 4. DEVI of the PLV-based scalp FC matrices of 9 preprocessing states from that of CE. The color in each box corresponds to the 9 preprocessing states in the horizontal line. MIP: more improper processing; MEP: more excessive preprocessing. A: delta band; B: theta band; C: alpha band; D: beta band.*

### 3.3.2 Deviation of the scalp EEG network metrics

The EEG networks were constructed by using the PLV. **Fig. 5** showed how the CPL, the CC, the GE, and the LE varied with the EEG quality. Seen from the general trends in the four subfigures, the CPL decreased, while the CC, the GE and the LE increased, with reduced contamination by the noise and loss of brain signals, that is, the EEG got more artifacts cleaned and even lost some components.

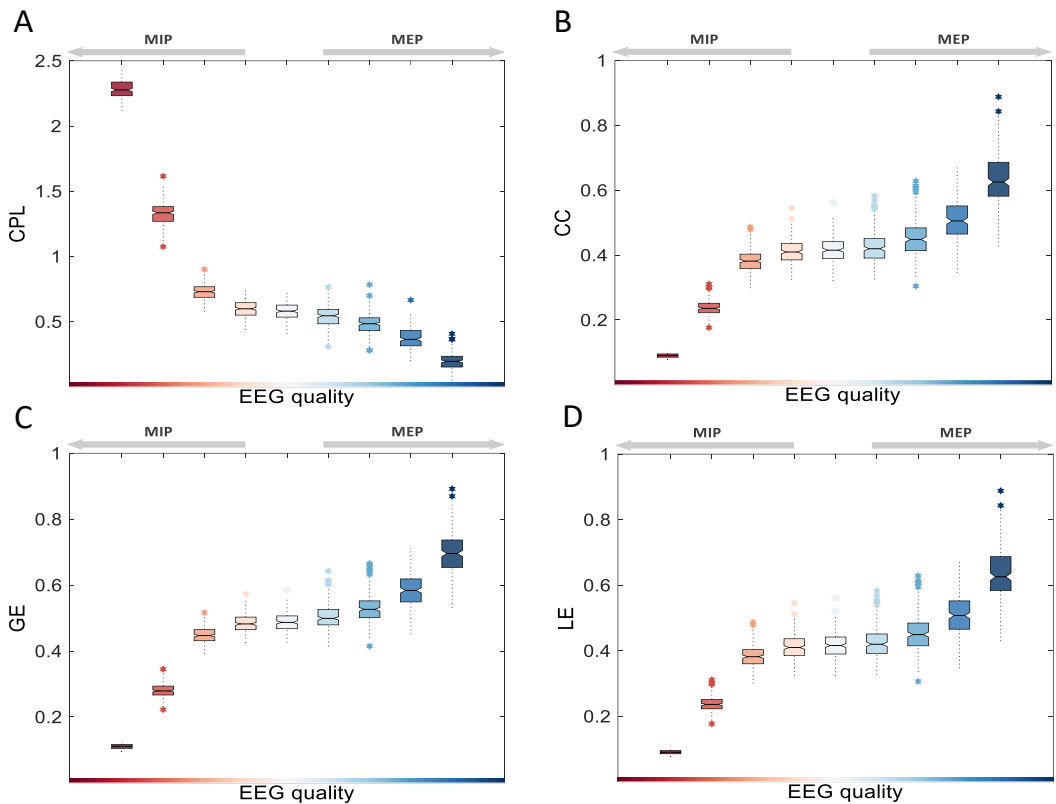

*Fig. 5. Varying trends of the graph theoretic metrics of the PLV-based scalp EEG networks with the preprocessing states. (A) CPL: characteristic path length, (B) CC: clustering coefficient, (C) GE: global efficiency, (D) LE: the local efficiency.*

To validate whether the results in the **Fig. 5** were affected by the selected coupling measure. The other 6 coupling measures were applied as well as the PLV.

The **Fig. 6** shows how the 4 graph theoretic network metrics, that is CPL, CC, GE, and LE, interacted with the 7 Coupling measures in the preprocessing states. The varying trends of network metric using one of the 7 tested measures were shown as the fold curves through the markers. As shown in **Fig. 6**A-D, the varying trends of 4 graph theoretic metrics presented a striking consistency varying with the EEG preprocessing states among the 7 coupling measures. From IPE to EPE, the CPL decreased, while the CC, the GE, and the LE increased. Notably, the DTF-PDC pair and the Coh-PLV pair nearly yielded the coincided curves, which can be inferred from the formulas of DTF and PDC, the formulas of COH and PLV. As commonly shown in the four subfigures, the nearly flat curves of iCOH may reveal that the iCOH based graph theoretic metrics were likely insensitive to both the contamination by the sensor noise and the loss of brain activities. Except for the CPL, the CC, GE, and LE commonly presented the trends of increasing from IPE to EPE. As to a particular preprocessing state, the CPL, GE, and LE across coupling measures followed the order that GC<PSI<iCOH<PDC<DTF<PLV<COH, while the CPL followed the inverse order.

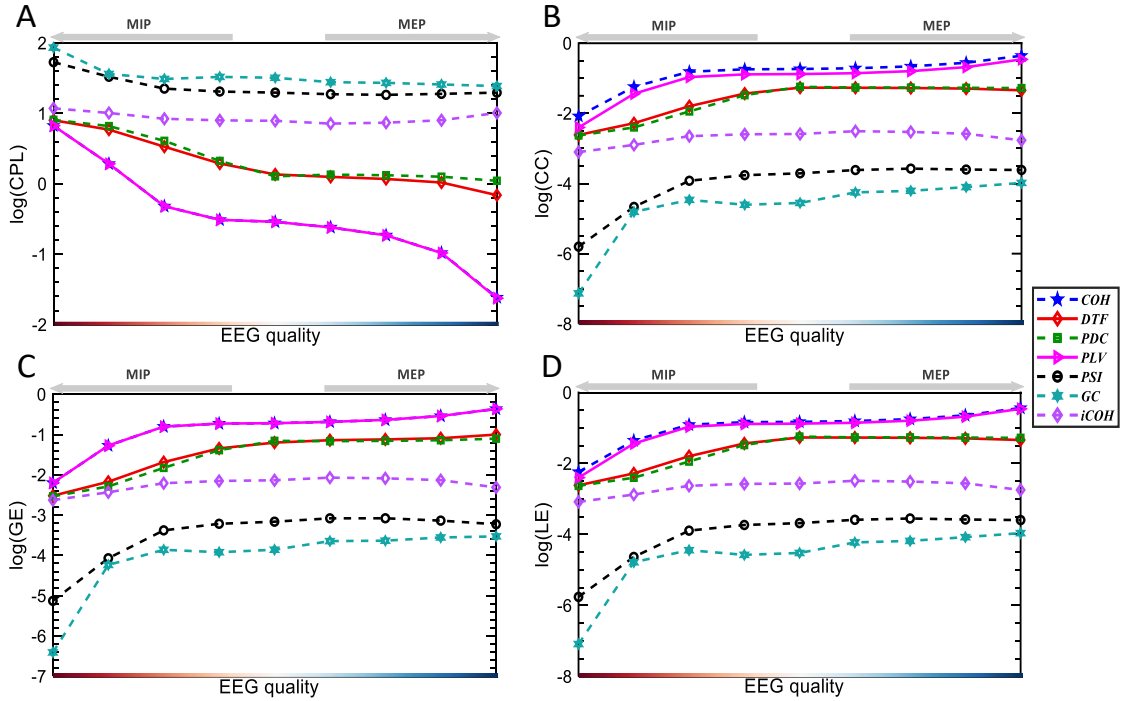

*Fig. 6. **The impacts of EEG preprocessing on the EEG network metrics and the interaction with functional connectivity measures.** **(A)**. CPL: characteristic path length, **(B)**. CC: clustering coefficient, **(C)**. GE: global efficiency, and **(D)**. LE: local efficiency.*

## 3.4 Impacts on the source imaging

The source localization was performed under different preprocessing states. The source activation intensity and the spatial dispersion (SD) were visualized and quantified. **Fig. 7**A shows the comparison of source imaging under the 9 preprocessing states in each frequency band. The central column was the source tomography estimated from the CE. For the IPE, they were set with the SNRs, 5, 8, 11, 14 dB from the leftmost to the middle column; while for the EPE, they were set with the SDRs, 20%, 40%, 55% and 70%, from the middle to the rightmost column. The SNRs and the SDRs were marked with color intensities, that is, the redder the bar, the smaller the SNR, the colder the bar, the greater the SDR. The colormap displays the intensities of the source activation. Compared with CE, the IPE tomography distributed more widely and was brighter with the SNR decreasing, while the EPE tomography distributed more focally and was lighter with the SDR increasing.

**Fig. 7**B provides an overview of the SDs under the 9 preprocessing states. Compared with CE, the SD values of both IPE and EPE decreased with the SNRs decreasing and the SDRs increasing. As to a frequency band, it was easily found that the bar heights in cold color changed greatly than that in warm color. This indicated that EPE may have greater impacts on the source dispersion than the IPE. Or say, the source dispersion is more sensitive to the loss of brain signals than to the contamination of sensor noise.

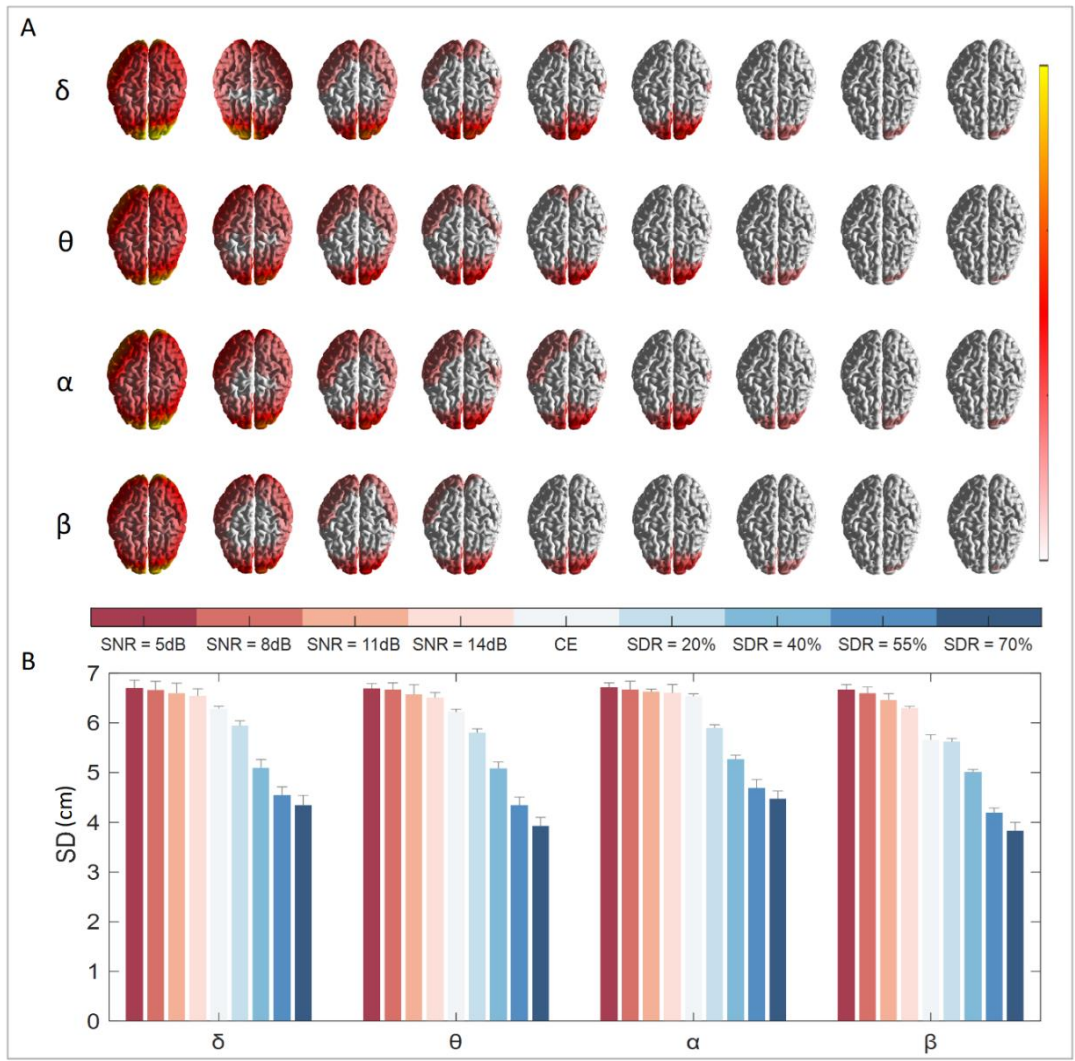

*Fig. 7. Impacts of preprocessed EEG quality on the source imaging.* (A). Tomography distribution of source intensities per state per frequency band; (B). The spatial dispersion (SD) of activated source distribution.

## 3.5 Association analysis

The **Fig. 8**A displayed how the PaLOSi values of the synthesized 200 EEG time series varied with the 9 preprocessing states. It is easily inspected that the PaLOSi increased exponentially in the IPE stage with relatively small SNRs, then increased slowly in the nearly "clean" stage around CE, and later increased exponentially in the EPE stage with relatively greater SDRs. The phenomenon that the IPE PaLOSi was smaller, and the EPE PaLOSi was larger than that of CE means that the PaLOSi positively correlated with the extent of EEG preprocessing. The PaLOSi reduced with more noise retained with more insufficient preprocessing, and the PaLOSi increased with more brain activities lost with more excessive preprocessing.

The normalized absolute deviations in the temporal, the frequency, and the spatial analysis were depicted using the pie charts in the **Fig. 8**B-D, respectively. In **Fig. 8**B-D, from left to right, the pie charts represent IPE (with signal-to-noise ratios of 5dB, 8dB, 11dB, 14dB), CE, and EPE (with brain signal loss ratios of 20%, 40%, 55%, 70%). The changes of PaLOSi were appended on the bottom of the B-D panels for the visual association analysis. Varying with the 9 preprocessing states, the relative normalized deviations of the spatial, spectral, and temporal analysis from CE generally followed the same trends as that of PaLOSi. Seen across the rows in each panel of the **Fig. 8**B-D, the spectral measures displayed great consistency, and the spatial measures showed relatively good consistency, but the temporal measures presented not good consistency. Or say, the changes across measures in the temporal and spatial domains are more obvious than that in the spectral analysis.

Additionally, in the results of time analysis, frequency analysis, and spatial analysis, it can be seen that the impact on postprocessing is minimal when the signal-to-noise ratio is 14dB, while it relatively increases when the brain signal loss ratio is 20%. This indicates that the harm caused by the loss of brain signals is more significant, suggesting a preference towards the left side rather than the right side in the preprocessing process.

Although the PaLOSi may not accurately quantify the deviations caused by the improper preprocessing, the PaLOSi is well associated with the deviations of all the investigated measures. In addition, it is easily found from the **Fig. 8** that when the PaLOSi belongs to the set [0.4, 0.6], the preprocessed EEG quality is closer to the CE and have less impacts on the postprocessing analysis. This may suggest that the PaLOSi of the EEG with acceptable quality should lay in an intermediate subset of [0, 1] but not approach the lower and upper borders. The associative relationship that PaLOSi correlated with the deviation of the temporal, frequency, and spatial outcomes can support the PaLOSi to be a promising EEG indicator.

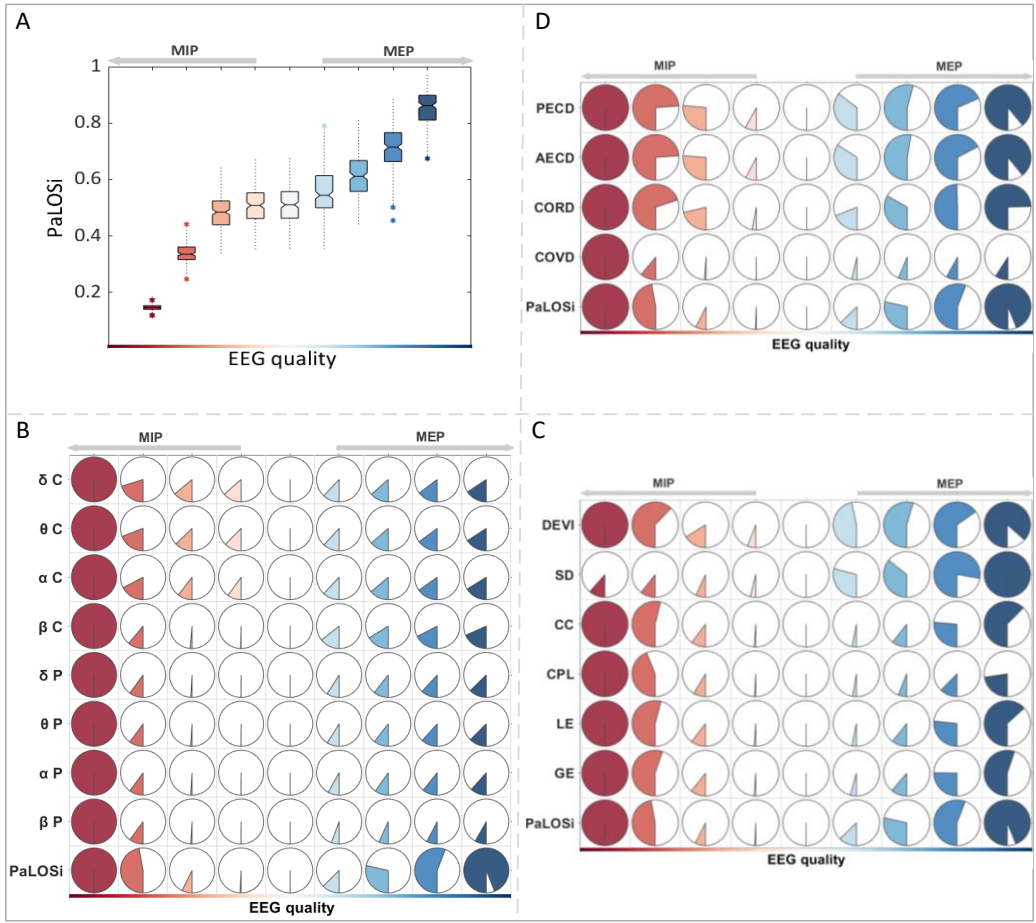

*Fig. 8. The association analysis within the preprocessing states, the normalized postprocessing deviations, and the PaLOSi. A. the PaLOSi varies with the EEG preprocessing states, B. The spectral deviations as to power spectra (P) and cross spectra (C) under the delta, theta, alpha, and beta frequency bands; C. The spatial deviations as to the scalp EEG network metrics and the source dispersion; D. The temporal deviations as to the cross-channel covariance (COV), the cross correlation (COR), the amplitude envelope correlation (AEC), and the power envelope correlation (PEC). Note that (1) the colors used in the pie charts correspond to the 9 preprocessing states; (2) all the investigated postprocessing deviations were normalized to the range [0,1].*

## 4. Discussion

### 4.1 The summary of the current study

To disclose the ambiguity in the quality control of MEEG studies, we performed a simulative study on how the preprocessed EEG quality affects the downstream postprocessing outcomes. The EEGs under the 9 preprocessing states were synthesized as the insufficiently preprocessed EEG (IPE) equipped with different SNRs and the excessively preprocessed EEG (EPE) by discarding the independent components with sorted explain variances. The downstream postprocessing involved: (1) the temporal analysis as to the cross-channel correlation, covariance, the amplitude envelope correlation, and the power envelope correlation; (2) the spectral analysis as to the power spectra and the cross spectra; (3) the spatial analysis as to the source dispersion and the fidelity of scalp EEG network using the graph theoretic metrics and the graph distance. The loss of postprocessing was evaluated by the deviation of the involved metrics. Lastly, the associative analysis was performed by summarizing the normalized absolute deviations and evaluating the interactive effects with our proposed metric PaLOSi. The results showed that the investigated postprocessing were greatly manipulated by the preprocessing states. And all the deviations of postprocessing were well associated with PaLOSi, which supports PaLOSi as a potential applicable quality control indicator.

## 4.2 The temporal deviations

As shown in the **Fig. 2**, the covariance did not show a significant change due to the preprocessing states. Or say, the covariance was not sensitive to the preprocessing states. Theoretically, the covariance measured the linear dependency of the time series between any two channels without standardization. Either the injection of sensor noise or the removal of the brain activities has no large impacts on the covariance. The correlation is just the covariance dividing by the standard deviations of the two random variables. This may indicate that the preprocessing states have substantial effects on the product of standard deviations. The COR, AEC and PEC were all sensitive to the preprocessing states. Both AEC and PEC were consistently sensitive to injection of sensor noise and removal of brain activity. For the COR, the EPE had a bit less effects than the IPE. The temporal deviation analysis may imply that the use of the temporal coupling measures such as the COR, AEC and PEC requires the high EEG quality and therefore the careful EEG preprocessing.

## 4.3 The frequency deviations

To our concern, frequency analysis is the main postprocessing method in the quantitative EEG analysis. Here, the deviations as to the power spectral density and the cross spectral were analyzed. Kaiser et al. (Kaiser et al. 2021) tested delta, theta, alpha, and beta frequency bands activity over the Fz, Cz, and Pz electrodes under the eyes-open and the eyes-closed conditions and demonstrated that poor-quality data with noise (IPE) significantly increased spectral power, which is in agreement with our study. More detailed, we investigated the effect of poor-quality data of brain signal loss (EPE) on spectral power and found that the loss of brain signals decayed the spectral power. The smaller the frequency, the greater the loss of spectral power (**Fig. 3A**). Moreover, under the case of IPE, the deviation between the power of all electrodes and CE is almost linear. The more IPE or the more EPE, the greater the spectral power of improper preprocessing deviated from that of CE.

## 4.4 The spatial deviations

EEG brain network was constructed using several classical Coupling measures and network metrics. The investigated network metrics were the CPL, the CC, and the GE and LE. The Coupling measures were COH, iCOH, DTF, GC, PDC, PLV, PSI, and DEVI. On the one hand, from IPE to CE to EPE, the DEVI firstly decreased and then increased as shown in **Fig. 4**. This may suggest that the more inappropriate the preprocessing, the greater the degree of brain network deviated. However, it is evident that the loss of brain signals leads to a more pronounced deviation in the network. On the other hand, the CPL decreased while the CC and the GE/LE increased with the extent of preprocessing as shown in **Fig. 7**.

Network isolation refers to the capability of specialized processing within closely interconnected clusters of nodes, which can be quantitatively measured using metrics such as CC and LE; network integration denotes the ability to combine information from distant nodes within a network, which can be quantitatively measured using metrics such as GE and CPL (Mheich, Wendling, and Hassan 2020; Rubinov and Sporns 2010; Meng and Xiang 2018). A lower CPL and higher GE indicate the ease of information flow and a faster parallel transfer of information in a network, becoming a superior integration of information. This suggests that, taking CE as the benchmark, IPE is less capable of exchanging information, less separated and less integrated in networks, while EPE shows the opposite trend. Even more, as the degree of EEG preprocessing increases, it becomes more integrated and less separated.

The small-worldness is described as the ratio of clustering coefficient to characteristic path length. Generally, a small-world network should possess both high integration and high segregation (Rubinov and Sporns 2010). With an increase in small-worldness, the capacity for functional integration and segregation also increases. In **Fig. 7**, clearly, noise enhances the heterogeneity of IPE, leading to lower small-worldness. Paradoxically, in the context of loss of brain signals, the small-worldness of EPE increases, seemingly contradicting common sense. Specifically, the small-world network in the brain should achieve an optimal balance between functional integration and segregation (Bassett and Bullmore 2006), rather than simply increasing.

The source estimation using eLoreta was applied on the synthesized 9 states of preprocessed EEGs. The source distribution in the showed that IPE had more sources activated with greater intensity, while EPE had fewer sources activated with weaker intensity, that is, both the intensity and the dispersion of brain sources reduced from IPE to EPE. From scalp EEG to source activity, in simple terms, involves a spatial filtering process. In comparison to CE, IPE contains additional noise, introducing some false sources on the cortex, while EPE causes genuine sources to vanish.

## 4.5 The associations with PaLOSi

This paper introduces a novel method—PaLOSi, which measures the homogeneity of cross-spectra. A higher PaLOSi value indicates more isomorphism in the spectra, while a lower value suggests greater heterogeneity. We anticipate that it could serve as one of the indicators for quality control in the frequency domain of EEG.

**Fig. 8**A demonstrates that, in comparison to CE, IPE has a lower PaLOSi value, indicating increased heterogeneity in spectral information due to the presence of noise. Conversely, EPE exhibits a higher PaLOSi value, suggesting enhanced isomorphism in the spectra. This could be attributed to increased dependency among the recombined signals in EPE after removing components obtained through the ICA method. In **Fig. 8**B-D, it can be observed that our proposed PaLOSi exhibits consistency with temporal, spectral, and spatial metrics during the preprocessing. This suggests that PaLOSi could serve as a frequency domain-based quality control measure. From the existing results, PaLOSi values are expected to be optimal within the range of 0.4 to 0.6. However, this range is only a preliminary assessment and intuitive understanding, requiring further investigation.

The temporal analysis, the spectral analysis, source imaging and brain network results suggest that the EEG with interference information or distortion may have a series of serious impacts for the consequent statistical analysis, biomarker identification, and interpretation as to brain mechanism accounting for the cognitive and behavior variables. In particular, the impact of excessive preprocessing (cooler colors in **Fig. 8**B-D) is more pronounced than insufficient preprocessing (warmer colors in **Fig. 8**B-D). However, when the signal-to-noise ratio is 14, the impact is not significant, but it becomes substantial with a 20% signal loss. This strongly reminds us not to engage in too much preprocessing (Delorme 2023) or, in other words, refrain from intensive preprocessing to achieve a high signal-to-noise ratio, as it may lead to signal loss, ultimately resulting in unfavorable trade-offs.

## 4.6 Outlook and limitation

The outlook and the limitation are as follows:
(1) This simulation study focusing on EEG can be extended to the other electrophysiological signals such as MEG. Although the CE state for the recorded EEG is never known, the results obtained in this study can be validated from the real analysis with convergent evidence.
(2) Neither the source functional connectivity analysis nor the impacts of preprocessing to the postprocessing in the source space were performed. The EEG scalp network is still used in the studies of recognition and classification. However, it is worth noting that the sensor spatial connectivity analysis is limited to the volume conduction effect and thus lacks enough neurophysiological interpretability.
(3) Besides, the connectivity analyses performed at the scalp level based on the Coupling measures derived from the MVAR model such as DTF, PDC, GC, etc. do not allow for interpretation in terms of anatomical interaction sources (Haufe et al. 2013; Van de Steen et al. 2019). Here, the interpretation is as to the scalp brain network.
(4) Note that the denoising of event related potentials (ERP) is grand averaging the time locked epochs and subtracting the pre-stimulus activities, being considerably different from the resting EEG artifact removal. Thus, the topic throughout this study is limited to the resting EEG.

# 5. Conclusion

To our best knowledge, it is the first study to fully investigate the impacts of preprocessing on the quantitative EEG analysis, mainly in terms of the spectral profiles and brain network analysis. We have demonstrated that the preprocessed data quality is crucial for the typical EEG features. We hope it can arouse the attention of the EEG community to large-scale EEG preprocessing and quality assurance.